\documentclass[a4paper,12pt]{article}
\usepackage{amssymb,graphicx}

\makeatletter
\renewcommand\section{\@startsection{section}{1}{\z@}                                    {-3.25ex\@plus -1ex \@minus -.2ex}                                    {1.5ex \@plus .2ex}                                    {\reset@font\Large\bfseries}}
\renewcommand\subsection{\@startsection{subsection}{2}{\z@}                                    {3.25ex \@plus 1ex \@minus.2ex}                                    {-1em}                                    {\reset@font\large\bfseries}}
\@addtoreset{equation}{section}
\makeatother
\renewcommand{\theequation}{\thesection.\arabic{equation}}
\input{tcilatex}
\begin{document}

%
%

\setcounter{page}{0} \topmargin0pt \oddsidemargin5mm \renewcommand{%
\thefootnote}{\fnsymbol{footnote}} \newpage \setcounter{page}{0} 
\begin{titlepage}
\begin{flushright}
Berlin Sfb288 Preprint  \\
hep-th/0203130
\end{flushright}
\vspace{0.2cm}
\begin{center}
{\Large {\bf Finite temperature correlation functions } }

{\Large {\bf from form factors } }

\vspace{0.8cm}
{\large  O.A.~Castro-Alvaredo and  A.~Fring }

\vspace{0.2cm}
{Institut f\"ur Theoretische Physik, 
Freie Universit\"at Berlin,\\
Arnimallee 14, D-14195 Berlin, Germany }
\end{center}
\vspace{0.5cm}
 
\renewcommand{\thefootnote}{\arabic{footnote}}
\setcounter{footnote}{0}

\begin{abstract}
We investigate proposals of how the form factor approach to compute
correlation functions at zero temperature can be extended to finite
temperature. For the two-point correlation function we conclude that the suggestion 
to use the usual form factor expansion with the modification of introducing 
dressing functions of various kinds is only suitable for free theories. 
Dynamically interacting theories 
require a more severe change of the form factor program. 
\medskip
 
\end{abstract}
\vfill{ \hspace*{-9mm}
\begin{tabular}{l}
\rule{6 cm}{0.05 mm}\\
Olalla@physik.fu-berlin.de\\
Fring@physik.fu-berlin.de\\
\end{tabular}}
\end{titlepage}
\newpage 

\section{Introduction}

The computation of correlation functions is one of the central objectives in
quantum field theories. In general, this can only be achieved by means of
perturbation theory in the coupling constant. Nonetheless, in 1+1 space-time
dimensions many exact results are known, in particular at zero temperature.
At present, one of the most successful approaches in this direction is the
form factor program. Originally this method was developed to compute
correlation functions for massive models at zero temperature \cite{Kar,Smir}%
. More recently, it has also been demonstrated, that the approach can be
employed successfully for the computation of massless correlation functions 
\cite{massless} for vanishing temperature. However, in a realistic set-up of
a physical experiment one needs to know such functions in the finite
temperature regime. Computing for instance physical quantities from linear
response theory one requires the response function in form of the canonical
two-point correlation function at finite temperature 
\begin{equation}
\left\langle \Delta \mathcal{O}(x,t)\Delta \mathcal{O}^{\prime }(x^{\prime
},t^{\prime })\right\rangle _{T}\,,  \label{tt}
\end{equation}
where $\Delta \mathcal{O}(x,t)=\mathcal{O}(x,t)-\left\langle \mathcal{O}%
(x,t)\right\rangle _{T}$. For the static case, e.g., for the computation of
electric and magnetic susceptibilities this goes back to \cite{Kirk}. Taking
for instance the local operators $\mathcal{O}(x,t)$ and $\mathcal{O}^{\prime
}(x,t)$ to be the current $J(x,t)$, the dynamical response is needed to
compute the conductivity by means of the celebrated Kubo formula \cite{Kubo}.

In \cite{LLSS} a proposal was made to adapt the form factor approach to
finite temperature computations. In there it was demonstrated for some
operators of the Ising model that the modified approach indeed reproduces
the expected results for the temperature dependent correlation functions,
even when boundaries are included. In order to make the method meaningful
several technical assumptions were required to eliminate various infinities.
Since not all of them can be justified in an entirely rigorous fashion,
further evidence is desirable to support the working of the proposed
prescription, even for the Ising model. The proposal is very appealing,
since apart from the introduction of a dressing function, the main
structures of the (T=0)-form factor approach are perpetuated. It was argued
in \cite{LLSS} for the Ising model, that the dressing functions admit an
interpretation in terms of density distribution functions. This observation
was taken up in \cite{LM} and it was conjectured that the interpretation
should also hold for interacting theories. Some checks which support the
validity of this conjecture for the one-point function were presented in 
\cite{LM}. Shortly afterwards, doubts were raised in \cite{HS} on the
working of these formulae for the two-point function, albeit only a counter
example which required a chemical potential was provided. An additional
controversy arose thereafter about the nature of the ``dressing function''
(see equation (\ref{fd})) which has to be employed in the context of the 
one-point function. In \cite
{GD}, it was proposed to employ the on-shell free energies, rather than the
pseudo-energies obtainable from the thermodynamic Bethe ansatz (TBA) as
suggested in \cite{LM}. For the one-point function evidence was provided in 
\cite{AW} that the proposal in \cite{GD} appears to be incorrect. No 
explicit claims concerning the two-point function were made in \cite{GD}.

The main purpose of this manuscript is to contribute to this debate and
provide further evidence for the (non)-validity of the various proposals. In
regard to the importance of (\ref{tt}), we want to focus especially on the
study of the expressions for the two-point functions. So far the only few
explicit computations using the dressed form factor approach may be found in 
\cite{LLSS,HS}.

Our manuscript is organized as follows: In section 2 we outline the various
proposals made so far to evaluate temperature dependent correlation
functions in the massive as well as in the massless regime. We investigate
these proposals for two models: the complex free Fermion/Federbush model
(section 3) and the scaling Yang-Lee model (section 4). In section 5 we
state our conclusions. In the appendix we assemble various properties of
functions which occur throughout our computations.

\section{Temperature dependent correlation functions}

We start by providing a concise review of the main features of the
prescription to compute temperature dependent n-point correlation functions
by means of form factors. In general a temperature state is described by a
density matrix $\rho $ and the expectation value for observables is the
trace over the product of this matrix with the observables. Taking $\left|
\psi \right\rangle $ to be eigenstates of the Hamiltonian $H$ with
eigenvalues $E_{\psi }$, and as usual $\beta =1/kT$ \ with $k$ being
Boltzmann's constant and $T$ \ the absolute temperature, the temperature
dependent n-point function for the observables $\mathcal{O}_{1}\ldots 
\mathcal{O}_{n}$ \ is defined as 
\begin{equation}
\left\langle \mathcal{O}_{1}(x_{1},t_{1})\ldots \mathcal{O}%
_{n}(x_{n},t_{n})\right\rangle _{T}:=\frac{1}{Z}\sum_{\psi }e^{-\beta
E_{\psi }}\left\langle \psi \right| \mathcal{O}_{1}(x_{1},t_{1})\ldots 
\mathcal{O}_{n}(x_{n},t_{n})\left| \psi \right\rangle \,.  \label{corrT}
\end{equation}
As usual, this expression is normalized by dividing with the partition
function 
\begin{equation}
Z=\sum_{\psi }e^{-\beta E_{\psi }}\left\langle \psi |\psi \right\rangle \,,
\end{equation}
which ensures that the trace over the density matrix $\rho =e^{-\beta H}/Z$
is one. The central idea of the ``dressed'' form factor program is now, as
in the finite temperature case \cite{Kar,Smir}, to reduce the computation of
the expansion (\ref{corrT}) to the computation of form factors 
\begin{equation}
F_{\psi }^{\mathcal{O}}=\left\langle \mathcal{O}(0)|\psi \right\rangle \,.
\label{FF}
\end{equation}
This is achieved simply by the insertion of $(n-1)$ complete states $%
\sum_{\psi }\left| \psi \right\rangle \left\langle \psi \right| =1$.
Suppressing for compactness the space-time dependence of the operators, this
reads 
\begin{equation}
\left\langle \mathcal{O}_{1}\ldots \mathcal{O}_{n}\right\rangle _{T}=\frac{1%
}{Z}\sum_{\psi _{0}\ldots \psi _{n-1}}e^{-\beta E_{\psi _{0}}}\left\langle
\psi _{0}\right| \mathcal{O}_{1}\left| \psi _{1}\right\rangle \left\langle
\psi _{1}\right| \mathcal{O}_{2}\left| \psi _{2}\right\rangle \ldots
\left\langle \psi _{n-1}\right| \mathcal{O}_{n}\left| \psi _{0}\right\rangle
.  \label{corrt}
\end{equation}
Thereafter one needs a meaningful prescription to relate matrix elements of
the form $\left\langle \psi ^{\prime }\right| \mathcal{O}\left| \psi
\right\rangle $ to those where the vacuum is on the left, as in (\ref{FF}),
and a shift operator $\omega _{t,x}\mathcal{O}(x,t)=f(x,t)\mathcal{O}(0)$
which moves the operator to the origin. In the following we will take $%
x^{\mu }=(-ir,0),$ which implies the restriction $r<\beta $ in order to
ensure that $\exp (-(\beta +r)H)$ is a trace class operator.

Let us now specify the states $\left| \psi \right\rangle $ to be
multi-particle states of the form 
\begin{equation}
\left| \psi \right\rangle =|Z_{\mu _{1}}^{\dagger }(\theta _{1})Z_{\mu
_{2}}^{\dagger }(\theta _{2})\ldots Z_{\mu _{n}}^{\dagger }(\theta
_{n})\rangle \,,
\end{equation}
where the operators $Z_{\mu }^{\dagger }(\theta )$ are creation operators
for a particle of type $\mu $ as a function of the rapidity $\theta $. These
operators are assumed to satisfy the Faddeev-Zamolodchikov algebra \cite{FZ} 
$Z_{i}^{\dagger }(\theta _{1})Z_{j}^{\dagger }(\theta
_{2})=S_{ij}^{kl}(\theta _{12})Z_{k}^{\dagger }(\theta _{2})Z_{l}^{\dagger
}(\theta _{1})$ with $S$ being the scattering matrix depending on the
rapidity difference $\theta _{12}=\theta _{1}-\theta _{2}$. The prescription 
$\left\langle \psi ^{\prime }\right| \mathcal{O}\left| \psi \right\rangle
\rightarrow \left\langle \mathcal{O}|\psi \psi ^{\prime }\right\rangle $
then reads 
\begin{eqnarray}
&&\!\!\!\!\!\!\!\!\!\!\!\!\!\!\!\left\langle Z_{\mu _{1}}(\theta
_{1}^{\prime })\ldots Z_{\mu _{n}}(\theta _{n}^{\prime })\right| \mathcal{O}%
(x,t)|Z_{\mu _{n}}^{\dagger }(\theta _{n})\ldots Z_{\mu _{1}}^{\dagger
}(\theta _{1})\rangle =  \nonumber \\
&&\qquad \quad \sum_{\text{all contractions}}\langle \mathcal{O}(x,t)|Z_{\mu
_{n}}^{\dagger }(\theta _{n})\ldots Z_{\mu _{1}}^{\dagger }(\theta _{1})Z_{%
\bar{\mu}_{1}}(\theta _{1}^{\prime -})\ldots Z_{\bar{\mu}_{n}}(\theta
_{n}^{\prime -})\rangle   \label{ac}
\end{eqnarray}
with $\theta ^{-}=\theta -i\pi +i\epsilon $ and $\epsilon $ is an
infinitesimal quantity and $\bar{\mu}$ being the anti-particle of $\mu $.
After shifting the operator to the origin, the r.h.s. of (\ref{ac}) involves
the n-particle form factors, which we denote as 
\begin{equation}
F_{n}^{\mathcal{O}|\mu _{1}\ldots \mu _{n}}(\theta _{1},\ldots ,\theta
_{n})\equiv \langle \mathcal{O}(0)|\mathcal{\,}Z_{\mu _{1}}^{\dagger
}(\theta _{1})\ldots Z_{\mu _{n}}^{\dagger }(\theta _{n})\rangle \,.
\end{equation}
These form factors have to satisfy various properties \cite{Kar,Smir}, such
as Watson's equations 
\begin{eqnarray}
F_{n}^{\mathcal{O}|\ldots \mu _{i}\mu _{i+1}\ldots }(\ldots ,\theta
_{i},\theta _{i+1},\ldots ) &=&F_{n}^{\mathcal{O}|\ldots \mu _{i+1}\mu
_{i}\ldots }(\ldots ,\theta _{i+1},\theta _{i},\ldots )S_{\mu _{i}\mu
_{i+1}}(\theta _{i,i+1})\,,\qquad   \label{W1} \\
F_{n}^{\mathcal{O}|\mu _{1}\ldots \mu _{n}}(\theta _{1}+2\pi i,\ldots
,\theta _{n}) &=&\gamma _{\mu _{1}}^{\mathcal{O}}\,F_{n}^{\mathcal{O}|\mu
_{2}\ldots \mu _{n}\mu _{1}}(\theta _{2},\ldots ,\theta _{n},\theta
_{1})\,\,,  \label{W2}
\end{eqnarray}
Lorentz invariance 
\begin{equation}
F_{n}^{\mathcal{O}|\mu _{1}\ldots \mu _{n}}(\theta _{1},\ldots ,\theta
_{n})=e^{s\lambda }F_{n}^{\mathcal{O}|\mu _{1}\ldots \mu _{n}}(\theta
_{1}+\lambda ,\ldots ,\theta _{n}+\lambda )\,,  \label{lo}
\end{equation}
and the so-called kinematic residue equation 
\begin{equation}
\limfunc{Res}_{{\small \bar{\theta}}\rightarrow {\small \theta }_{0}}\!%
{\small F}_{n+2}^{\mathcal{O}|\bar{\mu}\mu \mu _{1}\ldots \mu _{n}}{\small (%
\bar{\theta}+i\pi ,\theta }_{0}{\small ,\theta }_{1}{\small \ldots \theta }%
_{n}{\small )}=i[1-{\small \gamma _{\mu }^{\mathcal{O}}}\!\prod_{l=1}^{n}S_{%
\mu \mu _{l}}(\theta _{0l})]{\small F}_{n}^{\mathcal{O}|\mu _{1}\ldots \mu
_{n}}{\small (\theta }_{1},{\small \ldots ,\theta }_{n}{\small ).}
\label{kin}
\end{equation}
In (\ref{lo}) $s$ is the Lorentz spin of $\mathcal{O}$ and $\lambda \in \Bbb{%
C}$ an arbitrary shift. In equations (\ref{W2}) and (\ref{kin}) $\gamma
_{\mu }^{\mathcal{O}}$ is the factor of local commutativity defined through
the equal time exchange relation of the local operator $\mathcal{O}(x)$ and
the field $\mathcal{O}_{\mu }(y)$ associated to the particle creation
operators $Z_{\mu }^{\dagger }(\theta )$, i.e., $\mathcal{O}_{\mu }(x)%
\mathcal{O}(y)=\gamma _{\mu }^{\mathcal{O}}\,\mathcal{O}(y)\,\mathcal{O}%
_{\mu }(x)\,$\ for\thinspace \thinspace $\ x^{1}>y^{1}$. It is the
singularity $\lim_{\epsilon \rightarrow 0}Z_{\mu }^{\dagger }(\theta )Z_{%
\bar{\mu}}(\theta ^{-})\rightarrow \infty $, implicit in (\ref{kin}), which
is the reason for the presence of the $\epsilon $ in (\ref{ac}). The
renormalization prescription which eliminates these divergencies is outlined
in \cite{Balog} (see also \cite{LLSS,LM}). Other types of singularities
arise from the contractions of terms like $Z_{\mu }(\theta )Z_{\mu
}^{\dagger }(\theta )Z_{\nu }(\theta ^{\prime })Z_{\nu }^{\dagger }(\theta
^{\prime })\rightarrow \delta ^{2}(0)$. As demonstrated explicitly in \cite
{LLSS} such terms are absorbed in the (re)-normalization factor $Z$. In this
way the one-point function \cite{Balog,LLSS,LM,HS,GD,AW} 
\begin{eqnarray}
\left\langle \mathcal{O}(r)\right\rangle _{T} &=&\frac{1}{Z}%
\sum_{n=0}^{\infty }\sum_{\mu _{1}\ldots \mu _{n}}\int \frac{d\theta
_{1}\ldots d\theta _{n}}{n!(2\pi )^{n}}\exp \left\{ -\beta \left[
\varepsilon _{\mu _{1}}(\theta _{1})+\ldots \varepsilon _{\mu _{n}}(\theta
_{n})\right] \right\} \quad \qquad   \nonumber \\
&&\times \left\langle Z_{\mu _{1}}(\theta _{1})\ldots Z_{\mu _{n}}(\theta
_{n})\right| \mathcal{O}(r)|Z_{\mu _{n}}^{\dagger }(\theta _{n})\ldots
Z_{\mu _{1}}^{\dagger }(\theta _{1})\rangle \quad r<\frac{1}{T},
\end{eqnarray}
becomes a meaningful expression. In many cases this function is an important
normalization factor, however, for the reasons mentioned in the introduction
we will focus our attention on the two-point function. It results to \cite
{LLSS} 
\begin{eqnarray}
\left\langle \mathcal{O}(r)\mathcal{O}^{\prime }(0)\right\rangle _{T}
&=&\sum_{n=1}^{\infty }\sum_{\mu _{1}\ldots \mu _{n}}\int \frac{d\theta
_{1}\ldots d\theta _{n}}{n!(2\pi )^{n}}\prod_{i=1}^{n}\left[ f_{\mu
_{i}}(\theta _{i},T)e^{-rT\varepsilon _{\mu _{i}}(\theta _{i},T)}\right]  
\nonumber \\
&&\times F_{n}^{\mathcal{O}|\mu _{1}\ldots \mu _{n}}(\theta _{1},\ldots
,\theta _{n})\left[ F_{n}^{\mathcal{O}^{\prime }|\mu _{1}\ldots \mu
_{n}}(\theta _{1},\ldots ,\theta _{n})\right] ^{*}\,\quad r<\frac{1}{T}%
.\quad \quad   \label{twop}
\end{eqnarray}
This formula requires several explanations and comments: We dropped here as
usual another infinity coming from $n=0$ in the infinite sum. The sum over
the $\mu $ extends over particles and holes. This is understood in the way
that to each particle type present at zero temperature one may associate a
hole, such that each term at zero temperature which is summed over $n$%
-particles is mapped into $2^{n}$-terms at finite tempearture. According to 
\cite{LLSS}, the form factors involving holes may be constructed from the
ones of particles by an $i\pi $-shift 
\begin{equation}
F_{n}^{\mathcal{O}|\mu _{1}\ldots \text{particle}\ldots \mu _{n}}(\theta
_{1}\ldots ,\theta _{i},\ldots \theta _{n})=F_{n}^{\mathcal{O}|\mu
_{1}\ldots \text{hole}\ldots \mu _{n}}(\theta _{1}\ldots ,\theta _{i}-i\pi
,\ldots \theta _{n})\,.
\end{equation}
The origin of the hole interpretation is thus the crossing of particles from
bra to ket by means of (\ref{ac}) together with the renormalization
prescription. As explained in detail in \cite{LLSS}, in comparison with the
one-point function there are additional singular terms emerging in the
two-point functions. In the expansion they occur in terms in which the
product of the two form factors associated to $\mathcal{O}$ and $\mathcal{O}%
^{\prime }$ involve a different amount of particles. These terms are just
dropped, which could be a possible source for the difficulties we encounter
below in interacting theories.

\noindent The functions $f_{i}(\theta ,T)$ are the so-called filling
fractions 
\begin{equation}
f_{i}(\theta ,T)=\frac{1}{1-S_{ii}(0)\exp \left[ -\varepsilon _{i}(\theta
,T)\right] }  \label{fd}
\end{equation}
involving the functions $\varepsilon _{i}(\theta ,T)$. For the
non-interacting case, treated in \cite{LLSS}, this function is taken to be
the on-shell energy divided by the temperature $\varepsilon _{i}(\theta
,T)=m_{i}/T\cosh \theta $. When extending the validity of these formulae to
the interacting case, one may speculate on the nature of $\varepsilon
_{i}(\theta ,T)$. In \cite{LM} it was proposed to interpret $\varepsilon
_{i}(\theta ,T)$ as the pseudo-energies, which may be determined by means of
the thermodynamic Bethe ansatz equation \cite{TBAZam} 
\begin{equation}
\varepsilon _{i}(\theta ,\hat{r})=\hat{r}\,m_{i}\cosh \theta
+\sum\nolimits_{j}[\varphi _{ij}*\ln f_{j}](\theta )\,\,.  \label{TBA}
\end{equation}
As common we denote here $\hat{r}=m/T$, $m_{l}\rightarrow m_{l}/m$, with $m$
being the mass of the lightest particle in the model. By $\left( f*g\right)
(\theta )$$:=1/(2\pi )\int d\theta ^{\prime }f(\theta -\theta ^{\prime
})g(\theta ^{\prime })$ we denote the convolution of two functions and $%
\varphi _{ij}(\theta )=-id\ln S_{ij}(\theta )/d\theta $. In contrast, when
specializing the statement in \cite{LM} to fermionic statistics, i.e., $%
S_{ii}(0)=-1$ , it was suggested therein to take instead the free on-shell
energies divided by the temperature also in the interacting case. More
generally, this means that the eigenvalue $E_{\psi }$ of the Hamiltonian in (%
\ref{corrt}) is either taken to be the free on-shell or the pseudo-energy.
In each case, the pseudo-energies of the holes are simply the negative of
the ones of the particles 
\begin{equation}
\varepsilon _{\text{hole}}(\theta ,T)=-\varepsilon _{\text{particle}}(\theta
,T)\,.  \label{ph}
\end{equation}
A simple but key property satisfied by the two-point function is the
Kubo-Martin-Schwinger (KMS)-condition \cite{KMS}. Assuming to have a time
evolution operator $\omega _{t}$ at disposal, it is easily obtained from the
trace properties of the temperature dependent correlation function 
\begin{equation}
\left\langle \omega _{t}\mathcal{OO}^{\prime }\right\rangle
_{T}=\left\langle \mathcal{O}^{\prime }\omega _{t+i\beta }\mathcal{O}%
\right\rangle _{T}\Leftrightarrow \!\left\langle \mathcal{O}(x,t)\mathcal{O}%
^{\prime }(x^{\prime },t^{\prime })\right\rangle _{T}=\!\left\langle 
\mathcal{O}^{\prime }(x^{\prime },t^{\prime })\mathcal{O}(x,t+i\beta
)\right\rangle _{T}.
\end{equation}
For more detailed discussion on this formula see e.g., \cite{Haag}. For the
choice $x^{\mu }=(-ir,0)$, as in (\ref{twop}), this condition reads 
\begin{equation}
\left\langle \omega _{t}\mathcal{OO}^{\prime }\right\rangle
_{T}=\left\langle \mathcal{O}^{\prime }\omega _{t-\beta }\mathcal{O}%
\right\rangle _{T}\quad \Leftrightarrow \quad \left\langle \mathcal{O}(r)%
\mathcal{O}^{\prime }(0)\right\rangle _{T}=\left\langle \mathcal{O}^{\prime
}(0)\mathcal{O}(r-\beta )\right\rangle _{T}\,.  \label{KMS}
\end{equation}
Noting that $\left\langle \mathcal{O}(r)\mathcal{O}^{\prime
}(0)\right\rangle _{T}=\left\langle \mathcal{O}(0)\mathcal{O}^{\prime
}(-r)\right\rangle _{T}$, it is clear that (\ref{twop}) indeed satisfies the
condition (\ref{KMS}), provided that the form factors obey 
\begin{eqnarray}
&&F_{n}^{\mathcal{O}|n\times \text{holes\thinspace \thinspace }m\times \text{%
particles}}(\theta _{1},\ldots ,\theta _{n+m})\left[ F_{n}^{\mathcal{O}%
^{\prime }|n\times \text{holes\thinspace \thinspace }m\times \text{particles}%
}(\theta _{1}\ldots ,\theta _{n+m})\right] ^{*}=  \nonumber \\
&&F_{n}^{\mathcal{O}|n\times \text{particles\thinspace \thinspace }m\times 
\text{holes}}(\theta _{1},\ldots ,\theta _{n+m})\left[ F_{n}^{\mathcal{O}%
^{\prime }|n\times \text{particles\thinspace \thinspace }m\times \text{holes}%
}(\theta _{1},\ldots ,\theta _{n+m})\right] ^{*}.\quad \qquad   \label{con}
\end{eqnarray}
However, from one of Watson's equations (\ref{W2}) and Lorentz invariance (%
\ref{lo}) we easily derive that in general one picks up a factor $\exp i\pi
(s^{\prime }-s)$ on the r.h.s. of (\ref{con}), where $s,s^{\prime }$ are the
Lorentz spins of $\mathcal{O}$ and $\mathcal{O}^{\prime }$, respectively.
Consequently, there is no problem with KMS when $(s^{\prime }-s)\in 2\Bbb{Z}$
like for instance when the two operators coincide. The latter case was
treated in \cite{LLSS,LM,HS,GD,AW}. Despite the fact that the KMS condition
puts some structural restrictions on the two-point functions, it is not
constraining enough in what more precise functional details concern, e.g.,
it can not shed any light on the controversy \cite{LM,GD,AW} about the
precise nature of the dressing function (\ref{fd}). However, it dictates the
two functions $\varepsilon _{\mu }(\theta ,T)$ appearing explicitly in (\ref
{twop}) and (\ref{fd}) to be identical.

To clarify further the working of (\ref{twop}) it would be highly desirable
to compute this functions for more explicit models. Unfortunately there are
not many temperature dependent two-point functions known from alternative
approaches which one could compare with in order to settle the issue.
Nonetheless, various limits are known which one can take as benchmarks.

\subsection{The massless limit, conformal correlation functions}

\indent \ \ 

\vspace{0.2cm}

\noindent In conformal field theory various methods have been developed to
compute correlation functions. At vanishing temperature the most celebrated
approach is the one which exploits the structure of the Virasoro algebra,
such that the correlation functions obey certain differential equations \cite
{BPZ}. To include the temperature is also fairly simple in this case. It is
achieved just by mapping the observables from the plane to the cylinder, $%
z\rightarrow \exp (2\pi T\vartheta )$, $\mathcal{O}(z)\rightarrow (2\pi
T)^{-\Delta _{\mathcal{O}}}e^{-2\pi T\vartheta \Delta _{\mathcal{O}}}%
\mathcal{O}(\vartheta )$ 
\begin{eqnarray}
&&\left\langle \mathcal{O}(\vartheta _{1},\bar{\vartheta}_{1})\mathcal{O}%
^{\prime }(\vartheta _{2},\bar{\vartheta}_{2})\right\rangle
_{T}=\left\langle \mathcal{O}(\vartheta _{1})\mathcal{O}^{\prime }(\vartheta
_{2})\right\rangle _{T}\left\langle \mathcal{O}(\bar{\vartheta}_{1})\mathcal{%
O}^{\prime }(\bar{\vartheta}_{2})\right\rangle _{T}\qquad \qquad  \nonumber
\\
&&\qquad \quad \quad =(2\pi T)^{\Delta _{\mathcal{O}}+\Delta _{\mathcal{O}%
^{\prime }}+\bar{\Delta}_{\mathcal{O}}+\bar{\Delta}_{\mathcal{O}^{\prime
}}}\,\left\langle \mathcal{O}(z_{1})\mathcal{O}^{\prime
}(z_{2})\right\rangle _{T=0}\left\langle \mathcal{O}(\bar{z}_{1})\mathcal{O}%
^{\prime }(\bar{z}_{2})\right\rangle _{T=0}.\qquad  \label{CFTT}
\end{eqnarray}
By construction (\ref{CFTT}) satisfies the KMS condition, provided $%
\left\langle \mathcal{O}(z_{1})\mathcal{O}^{\prime }(z_{2})\right\rangle
_{T=0}=\left\langle \mathcal{O}^{\prime }(z_{2})\mathcal{O}%
(z_{1})\right\rangle _{T=0}$. Alternatively one can get some further
information on this function by exploiting the KMS condition on one of the
holomorphic sectors by adopting a proposal made in \cite{Todo}. Ignoring
normal ordering one obtains 
\begin{equation}
\left\langle \mathcal{O}(\vartheta _{1})\mathcal{O}^{\prime }(\vartheta
_{2})\right\rangle _{T}=\sum_{n=-\infty }^{\infty }\sum_{k=-\infty }^{\infty
}\exp [-2\pi iT(\vartheta _{1}n+\vartheta _{2}k)]\left[ \mathcal{O}_{n},%
\mathcal{O}_{k}^{\prime }\right] \,.  \label{Tod}
\end{equation}
For this to hold we only need to assume that for $\mathcal{O}(\vartheta
_{1}) $ and $\mathcal{O}^{\prime }(\vartheta _{2})$ exist Fourier-Laurent
mode expansions of the form 
\begin{equation}
\mathcal{O}(\vartheta )=\sum\nolimits_{n=-\infty }^{\infty }\exp (-2\pi
niT\vartheta )\mathcal{O}_{n}\,.
\end{equation}
In addition one makes use of the fact that the time evolution is governed by 
$\omega _{t}\mathcal{O}=e^{itL_{0}}\mathcal{O}e^{-itL_{0}}$, with $L_{0}$
being the zero mode generator of the Virasoro algebra. If then furthermore
the commutator $[L_{0},\mathcal{O}_{n}]=-n\mathcal{O}_{n}$ holds (this is
true for instance for $\mathcal{O}_{n}=L_{n}$ the modes of the
energy-momentum tensor, $\mathcal{O}_{n}=J_{n}^{a}$ the modes of a Kac-Moody
current, $\mathcal{O}_{n}=\phi _{n}$ the modes of a primary field, $\mathcal{%
O}_{n}=G_{n}$ the modes of an $N=1$ supersymmetric field\footnote{%
In \cite{Buch} it was shown, however, that supersymmetry and temperature
seem to be incompatible concepts. Only the vacuum states admit the
implementation of supersymmetry.}), the relation (\ref{Tod}) is derived
immediately. To make contact with (\ref{CFTT}) one needs of course to
incorporate a proper normal ordering prescription.

\begin{center}
\includegraphics[width=11.2cm,height=7.77cm]{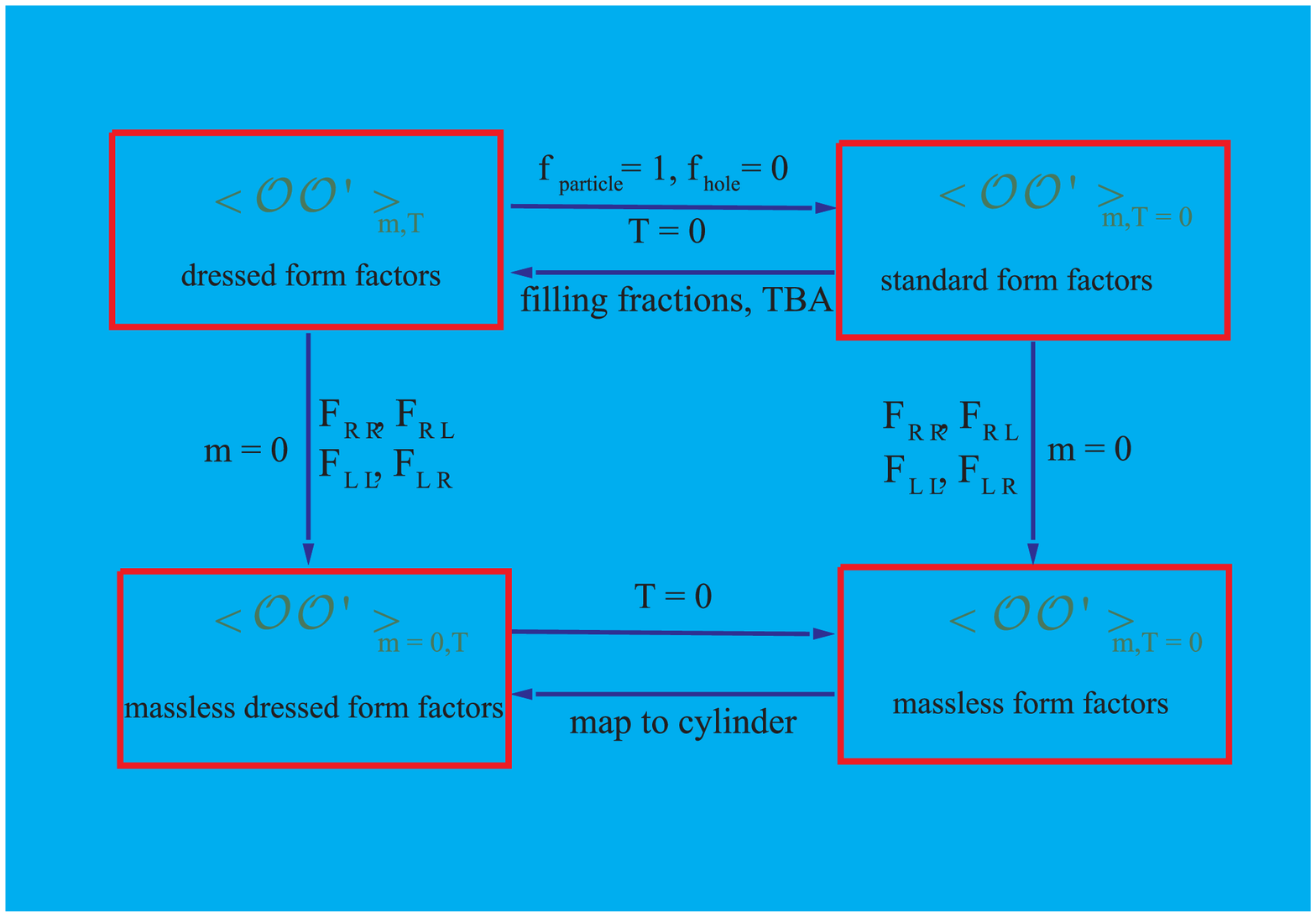}
\end{center}

\noindent {\small Figure 1: Two-point correlation functions in various
limits. } \medskip

Alternatively, we may compute the correlation functions by using the
(dressed) form factors related to the massless theory. The prescription of
taking the massless limit was originally introduced in \cite{triZam} within
the context of a scattering theory. It consists of replacing in every
rapidity dependent expression $\theta \rightarrow \theta \pm \sigma $, where
an additional auxiliary parameter $\sigma $ has been introduced. Thereafter
one should take the limit $\sigma \rightarrow \infty $, $m\rightarrow 0$
while keeping the quantity $\hat{m}=m/2\exp (\sigma )$ finite. For instance,
carrying out this prescription for the momentum yields $p_{\pm }=\pm \hat{m}%
\exp (\pm \theta )$, such that one may view the model as splitted into its
two chiral sectors and one can speak naturally of left (L) and right (R)
movers. Hence, having a function depending on the rapidities of $n$
particles, it will be mapped into $2^{n}$ related functions 
\begin{equation}
\lim_{\sigma \rightarrow \infty }f(\theta _{1}+\varkappa _{1}\sigma ,\ldots
,\theta _{n}+\varkappa _{n}\sigma )=f_{\nu _{1},\ldots ,\nu _{n}}(\theta
_{1},\ldots ,\theta _{n}),\left\{ 
\begin{array}{l}
\nu _{i}=L\text{ for }\varkappa _{i}=- \\ 
\nu _{i}=R\text{ for }\varkappa _{i}=+
\end{array}
\right. .  \label{massff}
\end{equation}

\noindent For the scattering matrix this means every massive amplitude is
duplicated $S(\theta )=$ $S_{LL}(\theta )=S_{RR}(\theta )$ and in addition
one obtains the amplitudes connecting the two chiral sectors $%
S_{RL/LR}(\theta )=\lim_{\sigma \rightarrow \infty }S(\theta \pm 2\sigma )$.

\noindent In \cite{massless} this prescription was also carried out for
expressions of form factors. In that case each n-particle form factor is
turned into $2^{n}$ n-particle form factors. Note that when considering (\ref
{massff}) for form factors this does in general not lead to the same
expressions as when taking the scattering matrices $S_{LL}(\theta
),S_{RR}(\theta ),S_{RL}(\theta ),S_{LR}(\theta )$ and determining the form
factors thereafter in the usual fashion. This two ways of carrying out the
limit only commute for form factors associated to operators whose Lorentz
spin is vanishing.

In order to be able to formulate the analogue of the expression (\ref{twop})
for the massless case one also requires the massless version of the dressing
function, i.e., the massless analogues of the pseudo-energies. For this one
can use once more the above recipe, such that the TBA-equations (\ref{TBA})
are replaced by the same equations with $S(\theta )\rightarrow S_{LL}(\theta
),S_{RR}(\theta ),S_{RL}(\theta ),S_{LR}(\theta )$ and $\hat{r}\,m_{i}\cosh
\theta \rightarrow \hat{r}\,\hat{m}_{i}\exp \theta $. The working of this
was confirmed in the analysis of \cite{triZam}.

Having now outlined the prescription to compute the massless (temperature
dependent) correlation functions, a non-trivial check is constituted by the
comparison with (\ref{CFTT}). A more constricting check is to start with (%
\ref{twop}) and carry out the massless and zero temperature limit. That is
checking the commutativity of the diagram in figure 1. In particular the
massless limit can possibly shed more light on the issue of different
dressing functions.

\section{The complex free Fermion/Federbush model}

We will now demonstrate the working of the previous approach with various
examples. Let us start with the complex free Fermion case (the complex
version of the Ising model) with Virasoro central charge $c=2\times 1/2$.
One reason for not considering directly the simpler case of self-conjugate
Fermions is that all formulae presented in this section also hold for the
more general Federbush model \cite{Feder,Fform}. This means we can also
regard the results of this section to hold for more exotic statistical
interaction. In general, the free Fermion is particularly attractive to
start with, since for many operators the higher n-particle form factors are
vanishing such that the infinite series in (\ref{twop}) terminates. For the
self-conjugate Fermion several two-point correlation functions at zero
temperature have been computed, e.g., in \cite{Hecht,DSC}. Despite the
simplicity of the model, only for very few operators massive temperature
dependent correlation functions have been evaluated by means of form factors 
\cite{LLSS}. In order to put formula (\ref{twop}) on firmer ground it is
therefore desirable to check its working first of all for a wider range of
operators.

Taking the operator $\mathcal{O}$ and $\mathcal{O}^{\prime }$ to be the
current, we will present all four cases illustrated in figure 1 in some
detail. The current-current correlation functions are particularly
interesting, since they occur explicitly in the application within the Kubo
formula. Adopting the notation of our recent exposition \cite{Fform}, we
consider the correlation function involving one of the chiral currents $%
J^{\pm }=J^{0}\pm J^{1}$. The only non-vanishing massive form factors for
these operators are 
\begin{equation}
F_{2}^{J^{\pm }|\bar{\imath}i}(\theta ,\tilde{\theta})=-F_{2}^{J^{\pm }|i%
\bar{\imath}}(\theta ,\tilde{\theta})=-i\pi me^{\mp \frac{\theta +\tilde{%
\theta}}{2}}\quad \,\,.  \label{J2}
\end{equation}
In the following we shall focus on the mutual correlator of $J\equiv J^{-}$.
Out of the four cases depicted in figure 1, the conformal case at zero
temperature is the easiest to treat and hence a good starting point.
According to the massless limit prescription (\ref{massff}), we obtain 
\begin{eqnarray}
F_{2,RR}^{J|\bar{\imath}i}(\theta ,\tilde{\theta}) &=&-F_{2,RR}^{J|i\bar{%
\imath}}(\theta ,\tilde{\theta})=-2\pi i\,\hat{m}\exp (\theta +\tilde{\theta}%
)/2\,,  \label{mf1} \\
F_{2,LL}^{J|\bar{\imath}i}(\theta ,\tilde{\theta}) &=&F_{2,LR}^{J|\bar{\imath%
}i}(\theta ,\tilde{\theta})=F_{2,RL}^{J|\bar{\imath}i}(\theta ,\tilde{\theta}%
)=0,  \label{mf2} \\
F_{2,LL}^{J|i\bar{\imath}}(\theta ,\tilde{\theta}) &=&F_{2,LR}^{J|i\bar{%
\imath}}(\theta ,\tilde{\theta})=F_{2,RL}^{J|i\bar{\imath}}(\theta ,\tilde{%
\theta})=0\,,  \label{:->}
\end{eqnarray}
such that (\ref{twop}) yields 
\begin{equation}
\left\langle J(r)J(0)\right\rangle _{m=0,T=0}=4\hat{m}^{2}\pi
^{2}\int_{-\infty }^{\infty }\frac{d\theta d\,\tilde{\theta}}{(2\pi )^{2}}%
\exp [-r\,\hat{m}(e^{\theta }+e^{\tilde{\theta}})](e^{\theta +\tilde{\theta}%
})=\frac{1}{r^{2}}\,.  \label{J0}
\end{equation}
Recalling that the current has scaling dimension $\Delta _{J}=1$, $\bar{%
\Delta}_{J}=0$, this agrees of course with the leading order term of the
well-known conformal $U(1)$-current-current two-point correlation function $%
k/r^{2}$ for level $k=1$, see e.g., \cite{book}. Note, that as it should be,
the auxiliary parameter $\hat{m}$ has vanished in the final expressions.
Raising now the temperature we can use the same expressions for the form
factors, but the proposal (\ref{twop}) dictates that we have to dress them
with the massless version of the filling fractions 
\begin{equation}
\hat{f}_{\pm }(\theta ,T)=\frac{1}{1+\exp (\mp \hat{m}/Te^{\theta })}\,.
\end{equation}
for particles ($+$) and holes ($-$). The values are in agreement with (\ref
{ph}). With this we compute 
\begin{eqnarray}
\left\langle J(r)J(0)\right\rangle _{m=0,T} &=&4\hat{m}^{2}\pi ^{2}\sum_{\mu
,\nu =\pm }\int\limits_{-\infty }^{\infty }\frac{d\theta d\,\tilde{\theta}}{%
(2\pi )^{2}}(e^{\theta +\tilde{\theta}})\hat{f}_{\mu }(\theta ,T)\hat{f}%
_{\nu }(\tilde{\theta},T)e^{-r\,\hat{m}(\mu e^{\theta }+\nu e^{\tilde{\theta}%
})}  \nonumber  \label{JT} \\
&=&\frac{\pi ^{2}T^{2}}{\sin ^{2}(\pi rT)}\,.  \label{JT2}
\end{eqnarray}
The result (\ref{JT2}) can of course also be obtained directly from the
mapping (\ref{CFTT}) and the correlation function at zero temperature (\ref
{J0}). Making now the model massive, we employ instead of (\ref{mf1}) and (%
\ref{mf2}) the form factors (\ref{J2}) and evaluate 
\begin{eqnarray}
\left\langle J(r)J(0)\right\rangle _{m,T=0} &=&m^{2}\pi
^{2}\int\limits_{-\infty }^{\infty }\frac{d\theta d\,\tilde{\theta}}{(2\pi
)^{2}}\exp [-rm(\cosh \theta +\cosh \tilde{\theta})](e^{\theta +\tilde{\theta%
}})  \nonumber \\
&=&m^{2}\left[ K_{1}(rm)\right] ^{2},  \label{JM}
\end{eqnarray}
where $K_{1}(x)$ is a modified Bessel function (see appendix). Using the
limiting behaviour (\ref{A2}), we recover as expected the conformal
correlation function (\ref{J0}) in the limit $m\rightarrow 0$. Considering
now the massive finite temperature regime, we have to include in the
previous computation the massive dressing function 
\begin{equation}
f_{\pm }(\theta ,T)=\frac{1}{1+\exp (\mp m/T\cosh \theta )}\,.
\end{equation}
Then we compute according to (\ref{twop}) 
\begin{eqnarray}
\!\!\!\!\!\!\!\!\left\langle J(r)J(0)\right\rangle _{m,T}\!\!\!
&=&\!\!m^{2}\pi ^{2}\sum_{\mu ,\nu =\pm }\int\limits_{-\infty }^{\infty }%
\frac{d\theta d\,\tilde{\theta}}{(2\pi )^{2}}(e^{\theta +\tilde{\theta}%
})f_{\mu }(\theta ,T)f_{\nu }(\tilde{\theta},T)e^{-r\,m(\mu \cosh \theta
+\nu \cosh \tilde{\theta})}  \nonumber \\
&=&\!\!m^{2}\!\left[ \hat{K}_{1}^{+}(m,r,T)\right] ^{2}\!\!.  \label{JMT}
\end{eqnarray}
The functions $\hat{K}_{\alpha }^{\pm }(m,r,T)$ are defined in the appendix.
The restriction on the arguments of the Bessel functions (\ref{Bessel})
reflects the conditions $r<1/T$ in (\ref{twop}). With the help of (\ref{KT0}%
) we recover the expression (\ref{JM}) for $T\rightarrow 0$. Taking instead
first the limit $m\rightarrow 0$ in (\ref{JMT}), we reproduce with (\ref{l+1}%
) the previously computed conformal correlator (\ref{JT2}). In conclusion
this means the different methods to compute $\left\langle
J(r)J(0)\right\rangle $ for several mass and temperature regimes are
consistent and indeed the diagram in figure 1 is commutative for the
considered choice of operators.

We proceed now similarly and compute the two-point correlation functions
involving various other operators. In what follows we will be less explicit
as for $\left\langle J(r)J(0)\right\rangle $ in the derivation of the finite
temperature and mass correlation function and just quote the final results.
Thereafter, we carry out the various limits by using the formulae quoted in
the appendix.

Recalling that the only non-zero form factors of the trace of the
energy-momentum tensor $\Theta $ are 
\begin{equation}
F_{2}^{\Theta |\bar{\imath}i}(\theta ,\tilde{\theta})=F_{2}^{\Theta |i\bar{%
\imath}}(\theta ,\tilde{\theta})=-2\pi im^{2}\sinh (\theta -\tilde{\theta}%
)/2,
\end{equation}
we obtain, according to (\ref{twop}), for its mutual correlation function

\begin{equation}
\left\langle \Theta (r)\Theta (0)\right\rangle _{T,m}=2m^{4}\left[ \hat{K}%
_{1}^{+}(m,r,T)^{2}-\hat{K}_{0}^{-}(m,r,T)^{2}\right] .
\end{equation}
We can verify the commutativity of the diagram in figure 1 similarly as for
the current-current correlator 
\begin{eqnarray}
\lim_{m\rightarrow 0}\left[ \lim_{T\rightarrow 0}\left\langle \Theta
(r)\Theta (0)\right\rangle _{T,m}\right] &=&\lim_{m\rightarrow 0}\left[
2m^{4}\left( K_{1}^{2}(rm)-K_{0}^{2}(rm)\right) \right] =0,  \label{TT1} \\
\lim_{T\rightarrow 0}\left[ \lim_{m\rightarrow 0}\left\langle \Theta
(r)\Theta (0)\right\rangle _{T,m}\right] &=&\lim_{T\rightarrow 0}\left[
0\right] =0\,.  \label{TT2}
\end{eqnarray}
The conformal limits in (\ref{TT1}) and (\ref{TT2}) reflect of course the
vanishing of the trace of the energy-momentum tensor. Noting that the energy
density operator $\epsilon $ with conformal dimension $\Delta _{\epsilon }=%
\bar{\Delta}_{\epsilon }=1/2$ is related to the trace as $\Theta =m\epsilon $%
, we obtain more interesting limits 
\begin{eqnarray}
\lim_{m\rightarrow 0}\left[ \lim_{T\rightarrow 0}\left\langle \epsilon
(r)\epsilon (0)\right\rangle _{T,m}\right] &=&\lim_{m\rightarrow 0}\left[
2m^{2}\left( K_{1}^{2}(rm)-K_{0}^{2}(rm)\right) \right] =\frac{2}{r^{2}}, \\
\lim_{T\rightarrow 0}\left[ \lim_{m\rightarrow 0}\left\langle \epsilon
(r)\epsilon (0)\right\rangle _{T,m}\right] &=&\lim_{T\rightarrow 0}\left[ 
\frac{2\pi ^{2}T^{2}}{\sin ^{2}(\pi rT)}\right] =\frac{2}{r^{2\Delta
_{\epsilon }+2\bar{\Delta}_{\epsilon }}},
\end{eqnarray}
which are again consistent. Recalling that for the (++)-component of the
energy momentum tensor $T^{++}\equiv \bar{T}$ \ with $\Delta _{\bar{T}}=2$, $%
\bar{\Delta}_{\bar{T}}=0$ the only non-vanishing form factors are 
\begin{equation}
F_{2}^{\bar{T}|\bar{\imath}i}(\theta ,\tilde{\theta})=F_{2}^{\bar{T}|i\bar{%
\imath}}(\theta ,\tilde{\theta})=\pi i/2m^{2}\exp (\theta +\tilde{\theta}%
)\sinh (\theta -\tilde{\theta})/2.
\end{equation}
The mutual correlation function results to 
\begin{equation}
\left\langle \bar{T}(r)\bar{T}(0)\right\rangle _{T,m}=\frac{m^{4}}{8}\left[ 
\hat{K}_{3}^{+}(m,r,T)\hat{K}_{1}^{+}(m,r,T)-\hat{K}_{2}^{-}(m,r,T)^{2}%
\right] \,.
\end{equation}
Once again the commutativity of the diagram in figure 1 is confirmed 
\begin{eqnarray}
\lim_{m\rightarrow 0}\left[ \lim_{T\rightarrow 0}\left\langle \bar{T}(r)\bar{%
T}(0)\right\rangle _{T,m}\right] \!\! &=&\!\!\lim_{m\rightarrow 0}\left[ 
\frac{m^{4}}{8}\left( K_{1}(rm)K_{3}(rm)-K_{2}^{2}(rm)\right) \right] =\frac{%
1}{2r^{4}},\,\,\quad \quad \\
\lim_{T\rightarrow 0}\left[ \lim_{m\rightarrow 0}\left\langle \bar{T}(r)\bar{%
T}(0)\right\rangle _{T,m}\right] \!\! &=&\!\!\lim_{T\rightarrow 0}\left[ 
\frac{1}{2}\frac{\pi ^{4}T^{4}}{\sin ^{4}(\pi rT)}\right] =\frac{c}{%
2r^{2\Delta _{\bar{T}}+2\bar{\Delta}_{\bar{T}}}}\,.
\end{eqnarray}
We also compute 
\begin{equation}
\left\langle \bar{T}(r)\Theta (0)\right\rangle _{T,m}=\frac{m^{4}}{2}\left[ 
\hat{K}_{1}^{-}(m,r,T)^{2}-\hat{K}_{0}^{+}(m,r,T)\hat{K}_{2}^{+}(m,r,T)%
\right] \,,
\end{equation}
together with the expected limiting behaviour 
\begin{eqnarray}
\lim_{m\rightarrow 0}\left[ \lim_{T\rightarrow 0}\left\langle \bar{T}%
(r)\Theta (0)\right\rangle _{T,m}\right] &=&\lim_{m\rightarrow 0}\left[ 
\frac{m^{4}}{2}\left( K_{1}^{2}(rm)-K_{0}(rm)K_{2}(rm)\right) \right]
=0,\quad \quad  \label{TT3} \\
\lim_{T\rightarrow 0}\left[ \lim_{m\rightarrow 0}\left\langle \bar{T}%
(r)\Theta (0)\right\rangle _{T,m}\right] &=&\lim_{T\rightarrow 0}\left[
0\right] =0\,.  \label{TT4}
\end{eqnarray}
Once again, the conformal limits in (\ref{TT3}) and (\ref{TT4}) reflect the
vanishing of the trace of the energy-momentum tensor. Replacing $\Theta
\rightarrow m\epsilon $ will only change in (\ref{TT3}) $m^{4}\rightarrow
m^{3}$ and the remaining limits are the expected ones.

\noindent However, there are also operators for which the prescription does
nor work so smoothly. As an example, we now want to compute $\left\langle 
\bar{T}(r)\mu (0)\right\rangle _{T,m}$, with $\mu $ being the disorder field
of the complex free Fermion theory. For this purpose, it is necessary first
to recall the expressions of the form factors related to this field, which
were computed in \cite{Fform}, and shown to be different from the ones
associated to the counterpart of this field in the Ising model. Similarly as
for the latter model, it was found that only the form factors involving an
even number of particles are non-zero. Since for $\bar{T}$ the only
non-vanishing form factors are the two-particle ones, the only non-vanishing
form factors of the field $\mu $ which will be relevant for these
computations are 
\begin{equation}
F_{2}^{\mu |\bar{\imath}i}(\theta ,\tilde{\theta})=-F_{2}^{\mu |i\bar{\imath}%
}(-\theta ,-\tilde{\theta})=i/2\left\langle \mu \right\rangle _{T=0}\exp
[(\theta -\tilde{\theta})/2]\cosh ^{-1}(\theta -\tilde{\theta})/2\,\,.
\end{equation}
With these data we compute, again according to (\ref{twop}), 
\begin{eqnarray}
\!\!\!\!\! &&\!\!\!\!\!\!\!\!\!\left\langle \bar{T}(r)\mu (0)\right\rangle
_{T,m}=\frac{T^{3}\,\left\langle \mu \right\rangle _{T=0}}{16}%
\int\limits_{r}^{1/T}dt\,e^{-2tm}\left[ \frac{2m}{T}\Phi \left( -e^{-\frac{m%
}{T}},\frac{1}{2},tT\right) \Phi \left( -e^{-\frac{m}{T}},\frac{3}{2}%
,tT\right) \right.   \nonumber \\
\!\!\!\!\! &&\!\!\!\!\!\!\!\!\!\left. -\Phi \left( -e^{-\frac{m}{T}},\frac{3%
}{2},tT\right) ^{2}+3\Phi \left( -e^{-\frac{m}{T}},\frac{1}{2},tT\right)
\Phi \left( -e^{-\frac{m}{T}},\frac{5}{2},tT\right) -t\rightarrow \frac{1}{T}%
-t\right] .\,\,\,\quad   \label{Tmu}
\end{eqnarray}
We employed Lerch's transcendental function $\Phi (x,s,\alpha )$ (see
appendix). In comparison with our previous computations we have one
integration remaining. This results from the fact that unlike before we have
now a term $(\cosh \tilde{\theta}+\cosh \theta )$ in the denominator, which
we eliminate by a differentiation with respect to $r$ and a subsequent
integration. Alternatively, we obtain the same result by a direct
computation using a variable substitution similar as in \cite{Hecht} $\sinh
\theta /2=r\cos \phi $, $\sinh \tilde{\theta}/2=r\sin \phi $ . By employing (%
\ref{needed}) the limits come out as 
\begin{equation}
\lim_{m\rightarrow 0}\left[ \lim_{T\rightarrow 0}\left\langle \bar{T}(r)\mu
(0)\right\rangle _{T,m}\right] =\lim_{m\rightarrow 0}\left[ \left\langle \mu
\right\rangle _{T=0}\frac{e^{-2rm}}{16r^{2}}\right] =\frac{\left\langle \mu
\right\rangle _{T=0}}{16r^{2}}\,.
\end{equation}
However, starting with the limit $m\rightarrow 0$ in (\ref{Tmu}) is
problematic, because the expression $\lim_{x\rightarrow -1}\Phi (x,s,\alpha )
$ for $\limfunc{Re}s<1$ is not well defined. We obtain similar phenomena for
other correlation functions involving different operators, such as $%
\left\langle J(r)\mu (0)\right\rangle _{T,m}.$

In principle one could extend this list of correlation functions involving
various other operators and support more and more our overall conclusion,
namely that the conjecture of a dressed form factor expansion (\ref{twop})
is meaningful for free theories even when the underlying statistics is
anyonic. Next we want to see whether the picture still remains the same for
dynamically interacting theories.

\section{The scaling Yang-Lee Model}

The scaling Yang-Lee model (or minimal $A_{2}^{(2)}$-affine Toda field
theory), like its conformal counterpart with Virasoro central charge $%
c=-22/5 $, is one of the simplest interacting integrable quantum field
theories in 1+1 space-time dimensions. It is an ideal starting point to test
general ideas, since it is comprised of only one massive particle which
couples to itself. This is reflected by the pole in the physical sheet of
its scattering matrix 
\begin{equation}
S_{YL}(\theta )=\frac{\sinh \theta +i\sin \pi /3}{\sinh \theta -i\sin \pi /3}
\label{SYL}
\end{equation}
which was proposed in \cite{CM}. Closed formulae for all n-particle form
factors for various components of the energy-momentum tensor were computed
in \cite{FFZam}. For our purposes we will just require the ones up to two
particles. We recall from \cite{FFZam} in a slightly different notation the
form factors associated to $T^{++}\equiv \bar{T}$%
\begin{eqnarray}
F_{0}^{\bar{T}} &=&-\frac{\pi m^{2}}{\sqrt{3}},\qquad \quad \qquad F_{1}^{%
\bar{T}}(\theta )=-\frac{i\pi m^{2}}{\nu 2^{5/2}3^{1/4}}e^{2\theta },\quad
\label{f1} \\
F_{2}^{\bar{T}}(\theta _{1},\theta _{2}) &=&\frac{\pi m^{2}}{8}\frac{\cosh
\theta _{12}-1}{\cosh \theta _{12}+1/2}e^{\theta _{1}+\theta _{2}}F_{\text{%
min}}(\theta _{12}),  \label{f2}
\end{eqnarray}
where the so-called minimal form factor is 
\begin{equation}
F_{\text{min}}(\theta )=\exp \left[ -8\int_{0}^{\infty }\frac{dt}{t}\frac{%
\sinh \frac{t}{2}\sinh \frac{t}{3}\sinh \frac{t}{6}}{\sinh ^{2}t}\sin
^{2}\left( \frac{t(i\pi -\theta )}{2\pi }\right) \right]
\end{equation}
and $\nu $ is a constant given by 
\begin{equation}
\nu =\exp \left( 2\int_{0}^{\infty }\frac{dt}{t}\frac{\sinh \frac{t}{2}\sinh 
\frac{t}{3}\sinh \frac{t}{6}}{\sinh ^{2}t}\right) =1.11154\ldots
\end{equation}
Our aim is to use these expressions and compute by means of the dressed form
factor formula (\ref{twop}) the two-point correlation functions.
Unfortunately, to our knowledge there exists no computation in the massive
and temperature dependent situation to compare with. However, in the
massless case we have two benchmarks, namely 
\begin{eqnarray}
\left\langle \bar{T}(r)\bar{T}(0)\right\rangle _{T=m=0} &=&-\frac{11}{5}%
\frac{1}{r^{4}},  \label{ex1} \\
\left\langle \bar{T}(r)\bar{T}(0)\right\rangle _{T,m=0} &=&-\frac{11}{5}%
\frac{\pi ^{4}T^{4}}{\sin ^{4}(\pi rT)}\,.  \label{ex2}
\end{eqnarray}
Here (\ref{ex1}) is just the well known two-point function from conformal
field theory $\left\langle \bar{T}(r)\bar{T}(0)\right\rangle
_{T=m=0}=c/2r^{-4}$, with $c=-22/5$, and (\ref{ex2}) is this formula mapped
to the cylinder according to (\ref{CFTT}) with $\Delta _{\bar{T}}=2$. Let us
therefore compute the massless scattering matrices and form factors from (%
\ref{SYL}) and (\ref{f1}), (\ref{f2}), respectively. According to the
prescription (\ref{massff}) outlined at the end of section 2, we compute 
\begin{equation}
S_{RR}(\theta )=S_{LL}(\theta )=S_{YL}(\theta ),\qquad S_{RL}(\theta
)=S_{LR}(\theta )=1  \label{SYLm}
\end{equation}
and 
\begin{eqnarray}
F_{0}^{\bar{T}} &=&F_{1,L}^{\bar{T}}=F_{2,LL}^{\bar{T}}=F_{2,LR}^{\bar{T}%
}=F_{2,RL}^{\bar{T}}=0 \\
F_{1,R}^{\bar{T}}(\theta ) &=&-\frac{i\pi \,\hat{m}^{2}}{\nu 2^{1/2}3^{1/4}}%
e^{2\theta },\qquad \\
F_{2,RR}^{\bar{T}}(\theta _{1},\theta _{2}) &=&\frac{\pi \,\hat{m}^{2}}{2}%
\frac{\cosh \theta _{12}-1}{\cosh \theta _{12}+1/2}e^{\theta _{1}+\theta
_{2}}F_{\text{min}}(\theta _{12}).
\end{eqnarray}
These expressions also exemplify our remark at the end of section 2, namely,
that one can not take the scattering matrices (\ref{SYLm}) and compute the
form factors thereafter from first principles. In that case we would obtain
an $L$ and $LL$ contribution, which are evidently vanishing when we carry
out the method directly on the formulae (\ref{f1}) and (\ref{f2}) for the
massive regime. The reason is simply that the ``massless prescription'' is
only a way to carry out the limit starting with the massive expressions, but
not a first principle concept. Nonetheless, viewing it in this sense it
works extremely well.

Let us commence with the zero temperature case. The one-particle
contribution can be computed analytically 
\begin{equation}
\left\langle \bar{T}(r)\bar{T}(0)\right\rangle _{T=m=0}^{(1)}=-\int \frac{%
d\theta }{2\pi }\left| F_{1,R}^{\bar{T}}(\theta )\right| ^{2}e^{-\hat{m}%
re^{\theta }}=-\frac{\pi \sqrt{3}}{2\nu ^{2}}\frac{1}{r^{4}}=-\frac{%
2.2020498\ldots }{r^{4}}\,.  \label{echtgeil}
\end{equation}
We observe, that the expected value is already almost saturated. To compute
the two particle contribution is a fairly simple numerical exercise. We
evaluate 
\begin{equation}
\left\langle \bar{T}(r)\bar{T}(0)\right\rangle _{T=m=0}^{(2)}=\int \frac{%
d\theta _{1}d\theta _{2}}{2(2\pi )^{2}}\left| F_{2,RR}^{\bar{T}}(\theta
_{1},\theta _{2})\right| ^{2}e^{-\hat{m}r(e^{\theta _{1}}+e^{\theta
_{2}})}\,,  \label{2}
\end{equation}
and present our results in table 1 \medskip

\begin{center}
\begin{tabular}{||c||c|c|c|c|c|c||}
\hline
$r$ & $0.0001$ & $0.0003$ & $0.0005$ & $0.001$ & $0.01$ & $0.1$ \\ \hline
$r^{4}\left\langle \bar{T}(r)\bar{T}(0)\right\rangle _{T=m=0}^{(2)}10^{3}$ & 
$0.5631$ & $1.9435$ & $2.0457$ & $2.0487$ & $2.0487$ & $2.0487$ \\ \hline
\end{tabular}
\end{center}

\medskip \noindent \noindent {\small Table 1: Two-particle contribution to $%
\left\langle \bar{T}(r)\bar{T}(0)\right\rangle _{T=m=0}$ . } \medskip

As mentioned in \cite{FFZam}, when summing up according to (\ref{twop}) with 
$f_{\mu }(\theta ,T)=1$, this contribution enters with a positive sign in
comparison with the one-particle contribution due to the non-unitarity of
the model. As expected, similar to the ($T=0$, $m\neq 0$)-case, carried out
in \cite{FFZam}, we observe an extremely fast convergence of the series
towards the expected value (\ref{ex1}). The numbers in table 1 confirm the
general observation, which was also made in \cite{FFZam}, that for extremely
small values of $r$ the higher particle contributions become more important.
Considering a regime for $r>0.001$, it will not be necessary to include also
the three-particle contribution in order to reach our main conclusion.
Nonetheless, in principle this could be done easily with some Monte Carlo
integration, just at the cost of longer computing time, since in \cite{FFZam}
all n-particle form factors were already provided.

Notice further that the parameter $\hat{m}$ plays no role anymore. As
observed before, in the analytical computation (\ref{echtgeil}) it cancels
explicitly. This phenomenon is less apparent in the two-particle formula (%
\ref{2}), but we convinced ourselves that $\hat{m}$ may be re-scaled without
altering the outcome of the numerical computation. Thus the expressions are
mass independent as they should be. In what follows we will use this fact to
simplify notations and scale $\hat{m}$ to one.

Let us now embark upon the non-zero temperature case. By means of the
conjecture (\ref{twop}), we have to compute for the one-particle
contribution 
\begin{eqnarray}
\left\langle \bar{T}(r)\bar{T}(0)\right\rangle _{T,m=0}^{(1)} &=&-\int \frac{%
d\theta }{2\pi }\left( f_{+}(\theta ,T)\left| F_{1,R}^{\bar{T}}(\theta
)\right| ^{2}e^{-rT\varepsilon (\theta ,T)}\right.  \nonumber \\
&&\left. +f_{-}(\theta ,T)\left| F_{1,R}^{\bar{T}\bar{T}}(\theta -i\pi
)\right| ^{2}e^{rT\varepsilon (\theta ,T)}\right) e^{4\theta }  \label{T1}
\end{eqnarray}
According to the conjectures in \cite{LM,AW} or if we extend the proposal 
in \cite{GD} to the two-point function, we can choose for
the $\varepsilon (\theta ,T)$ in (\ref{T1}) and the corresponding dressing
functions (\ref{fd}) either $\varepsilon _{\text{TBA}}(\theta ,T)$,
determined by the massless version of the TBA-equation (\ref{TBA}) or $%
\varepsilon _{\text{free}}(\theta ,T)=e^{\theta }/T$, respectively. Solving
first numerically the massless TBA-equation by means of a standard iteration
procedure, we can compare the two dressing functions $f_{\pm }(\varepsilon _{%
\text{TBA}}(\theta ,T))$ and $f_{\pm }(\varepsilon _{\text{free}}(\theta
,T)) $. Our results are depicted in figure 2.

\begin{center}
\includegraphics[width=11.2cm,height=7.77cm]{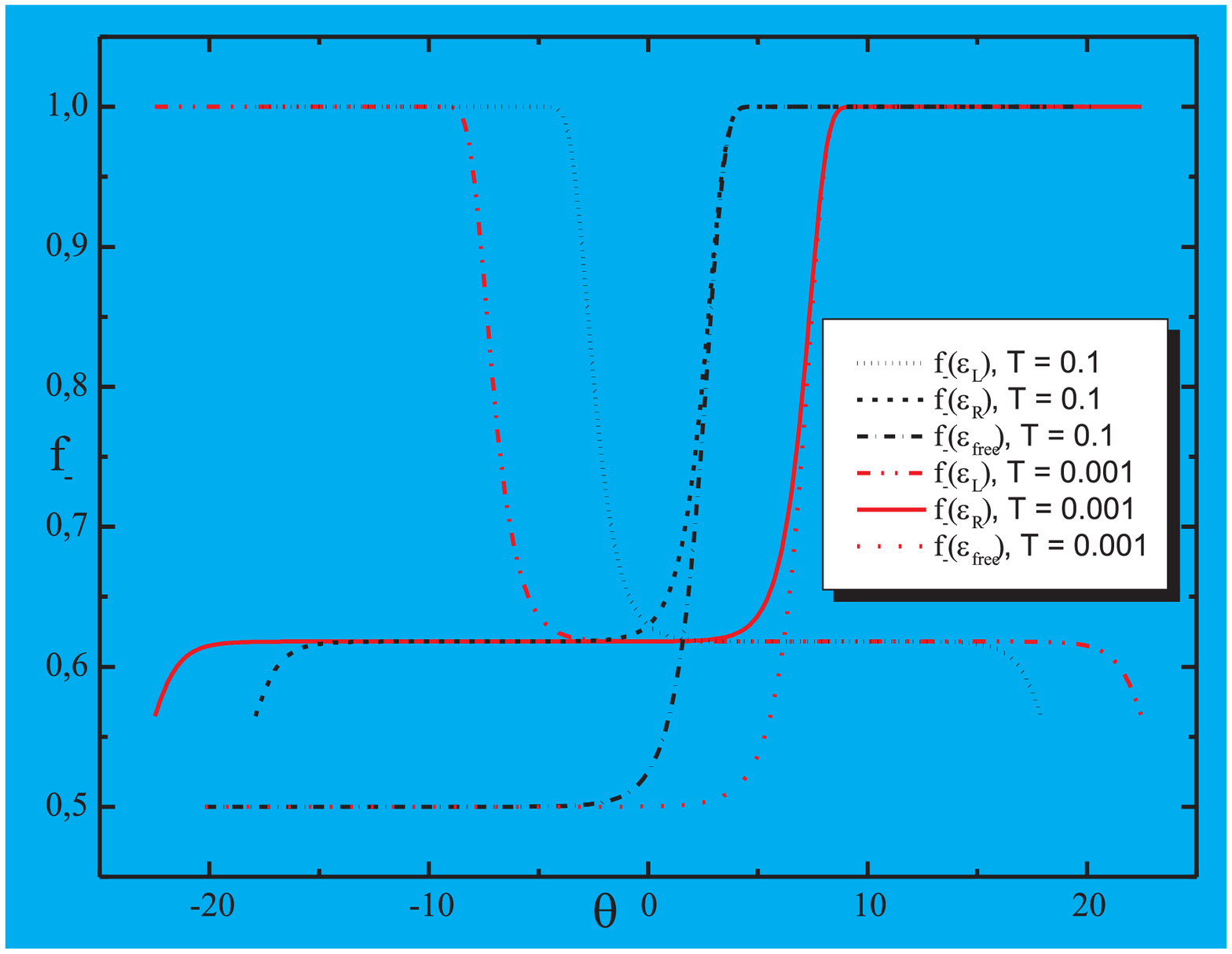}
\end{center}

\noindent {\small Figure 2: Two possible dressing functions }$f_{\pm
}(\varepsilon _{\text{TBA}}(\theta ,T))${\small \ and }$f_{\pm }(\varepsilon
_{\text{free}}(\theta ,T))${\small . } \medskip

We plotted the corresponding functions for the left and right movers in
order to exhibit the symmetry of the TBA solutions. The solutions for the
left movers are not important for what follows. We observe that $%
f_{-}(\varepsilon _{\text{TBA}}(\theta ,T))$ acquires plateaux at $%
f_{-}(\varepsilon _{\text{TBA}}(\theta ,T))=2/(1+\sqrt{5})$ which are
characteristic for all minimal affine Toda field theories. Furthermore, we
see the well-known fact that in the large rapidity regime the TBA solutions
merge with the free one. Nonetheless, the two functions $f_{\pm }^{\text{TBA}%
}(\theta ,T)$ and $f_{\pm }^{\text{free}}(\theta ,T)$ differ quite
substantially within a large rapidity regime. However, in the computation in
which they are actually needed, namely in (\ref{T1}), this regime is
negligible. This is essentially due to the factor $e^{4\theta }$. This means
the issue of controversy on the difference between the two functions $f_{\pm
}^{\text{TBA}}(\theta ,T)$ and $f_{\pm }^{\text{free}}(\theta ,T)$ is 
irrelevant if we would extend it to the context of the two-point functions. 
Let us therefore take $f_{\pm }^{\text{free}%
}(\theta ,T)$, for which we can compute (\ref{T1}) analytically 
\begin{equation}
\left\langle \bar{T}(r)\bar{T}(0)\right\rangle _{T,m=0}^{(1)}=-\frac{\pi 
\sqrt{3}}{2\nu ^{2}}T^{4}\left[ \Phi (-1,4,rT)+\Phi (-1,4,1-rT)\right] \,.
\label{111}
\end{equation}
Here $\Phi (x,s,\alpha )$ is again Lerch's transcendental function, which
was already encountered in section 3 (see also appendix). Extrapolating the
behaviour from the cases ($m\neq 0,T=0$) and ($T=0,m=0$), we expect that the
main contribution to the sum comes from the one-particle form factors. It is
therefore instructive to compare the ratio of the expected value (\ref{ex2})
and $\left\langle \bar{T}(r)\bar{T}(0)\right\rangle _{T\neq 0,m=0}^{(1)}$.
We depict this comparison in figure 3.

\begin{center}
\includegraphics[width=11.2cm,height=7.77cm]{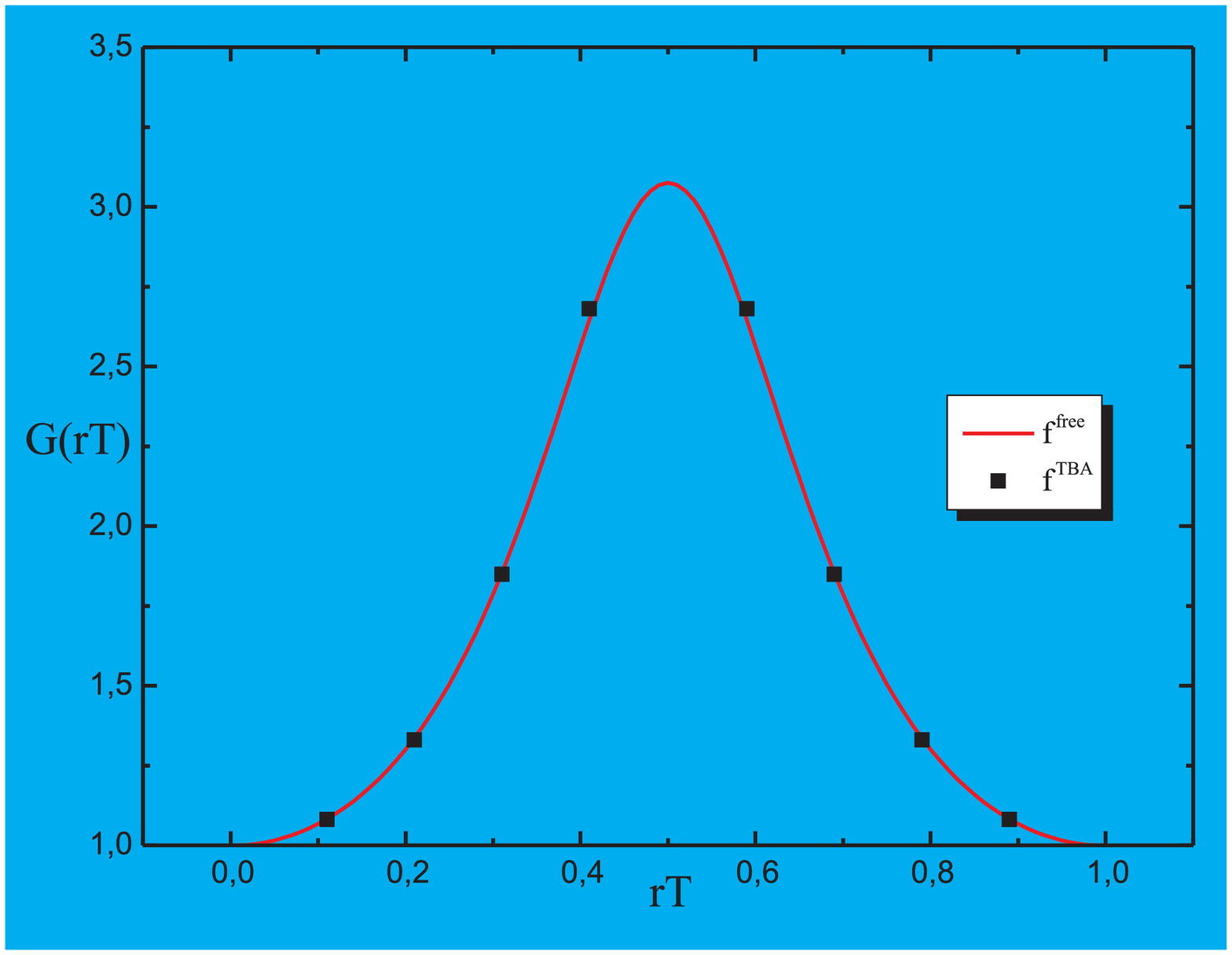}
\end{center}

\noindent {\small Figure 3: Exact correlation function \ versus dressed one
particle form factor contribution G(rT):=$\left\langle \bar{T}(r)\bar{T}%
(0)\right\rangle _{T\neq 0,m=0}/\left\langle \bar{T}(r)\bar{T}%
(0)\right\rangle _{T\neq 0,m=0}^{(1)}.$}

We observe a deviation of up to 300\%, which when recalling the excellent
agreement at this level of the cases ($m\neq 0,T=0$) and ($T=0,m=0$), sheds
a rather pessimistic light on the working of the conjecture (\ref{twop}) in
the interacting case. In fact, we only observe a reasonable match when $%
0<rT<0.01$ or $0.99<rT<1.0$, but this is unfortunately just a regime in
which we can approximate in (\ref{ex2}) $\sin ^{4}(\pi rT)\sim (\pi rT)^{4}$
such that the temperature effect becomes irrelevant. For the sceptical
reader we include in figure 3 also some points obtained numerically be using 
$f_{\pm }^{\text{TBA}}(\theta ,T)$ instead of $f_{\pm }^{\text{free}}(\theta
,T)$ in (\ref{T1}). The two different cases may hardly be distinguished.

Let us see whether the next order contributions can improve the situation.
For this we have to compute 
\begin{eqnarray}
\left\langle \bar{T}(r)\bar{T}(0)\right\rangle _{T,m=0}^{(2)}\!\!\!\!
&=&\!\!\!\!\int \frac{d\theta _{1}d\theta _{2}}{2(2\pi )^{2}}\left[
f_{+}(\theta _{1},T)f_{+}(\theta _{2},T)\left| F_{2,RR}^{\bar{T}}(\theta
_{1},\theta _{2})\right| ^{2}e^{-rT(\varepsilon (\theta _{1},T)+\varepsilon
(\theta _{2},T))}\right.   \nonumber \\
&&\!\!\!\!\!\!\!\!\!\!\!\!\!\!\!\!+f_{-}(\theta _{1},T)f_{-}(\theta
_{2},T)\left| F_{2,RR}^{\bar{T}}(\theta _{1},\theta _{2})\right|
^{2}e^{rT(\varepsilon (\theta _{1},T)+\varepsilon (\theta _{2},T))} 
\nonumber \\
&&\!\!\!\!\!\!\!\!\!\!\!\!\!\!\!\!\left. +2f_{+}(\theta _{1},T)f_{-}(\theta
_{2},T)\left| F_{2,RR}^{\bar{T}}(\theta _{1},\theta _{2}-i\pi )\right|
^{2}e^{-rT(\varepsilon (\theta _{1},T)-\varepsilon (\theta _{2},T))}\right] .
\label{tschüssi}
\end{eqnarray}
We have already all the ingredients to compute this apart from the
particle-hole form factor $F_{2,RR}^{\bar{T}}(\theta _{1},\theta _{2}+i\pi )$%
. We compute 
\begin{equation}
F_{2,RR}^{\bar{T}}(\theta _{1},\theta _{2}-i\pi )=-\frac{\pi \,}{2\nu ^{8}}%
\left( \frac{2\cosh \theta _{12}+1}{2\cosh \theta _{12}-1}\right) \frac{%
\tanh ^{2}(\theta _{12}/2)}{F_{\text{min}}(\theta _{12})}e^{\theta
_{1}+\theta _{2}}\,.
\end{equation}
Assembling all we find similar values in the entire range of $0<rT<1$%
\begin{equation}
\left\langle \bar{T}(r)\bar{T}(0)\right\rangle _{T,m=0}^{(2)}/\left\langle 
\bar{T}(r)\bar{T}(0)\right\rangle _{T,m=0}\sim 0.001\pm 4\times 10^{-4}\,,
\label{:-)}
\end{equation}
independently of the different choices for the dressing functions. Thus,
assuming that the convergence of the series in (\ref{twop}) does not change
radically when the temperature is switched on, the higher order n-particle
contributions will not rescue the proposal for this case.

\section{Conclusions}

We provided more evidence which supports the proposal of LeClair, Lesage,
Sachdev and Saleur \cite{LLSS} to use dressed form factors for the
computation of two-point correlation functions. The method seems to work
well for the free Fermion case and in addition for theories with anyonic
statistics. Concerning dynamically interacting theories we reach a similar
conclusion as drawn in \cite{HS} on the base of an example involving a
chemical potential: Namely that it fails to work. As a simple counter
example we have studied the scaling Yang-Lee model. This conclusion is
reached independently of the choices for the
dressing functions $f(\varepsilon _{%
\text{TBA}}(\theta ,T))$ or $f(\varepsilon _{\text{free}}(\theta
,T)) $.

Despite this slightly pessimistic result concerning the proposal in its
present form, it was shown in \cite{LLSS} that it also works for the
computation of correlation functions in the presence of boundaries, albeit
only for the Ising model. Based on this result one may conjecture that it
can also be successfully applied to defect systems \cite{OAnew}. In fact in
this context the only interesting integrable theories are those for which
the approach seems to work for the bulk theories, namely free theories,
possibly with anyonic statistical interaction. It was shown recently \cite
{OAF}, that these theories are the only bulk theories which allow a
simultaneous occurrence of reflection and transmission.

At present the challenge remains to show whether the form factor approach to
compute correlation functions can be extended successfully to the finite
temperature regime in complete generality, namely also for dynamically
interacting theories. In order to complete this task it would also be
interesting to compare with existing alternative approaches, e.g., \cite{SL}.

\medskip

\noindent \textbf{Acknowledgments: } We are grateful to the Deutsche
Forschungsgemeinschaft (Sfb288) and INTAS project 99-01459 
for financial support and to M. Karowski, A.
LeClair, I.T. Todorov for useful comments.

\setcounter{section}{0} \renewcommand{\thesection}{} \renewcommand{%
\theequation}{\Alph{section}.\arabic{equation}}

\section{Appendix}

In this appendix we assemble some properties of various functions which are
important for our computations. Some of them are standard whereas others are
specific to the present context. One of the most ubiquitous functions in
this context are the modified Bessel functions, whose integral
representations are given by 
\begin{equation}
K_{\alpha }(z)=\int_{0}^{\infty }dt\,\exp (-z\cosh t)\cosh \alpha t\,\quad 
\text{for\quad }\left| \arg z\right| <\frac{\pi }{2}\,\,.  \label{Bessel}
\end{equation}
We recall the well-known limiting behaviour 
\begin{equation}
\lim_{x\rightarrow 0}K_{\alpha }(x)\sim 2^{\alpha -1}\Gamma (\alpha
)x^{-\alpha }\quad \func{Re}\alpha >0,\qquad \quad \lim_{x\rightarrow
0}K_{0}(x)\sim -\ln x\,.  \label{A2}
\end{equation}
It is convenient to introduce the function 
\begin{equation}
\hat{K}_{\alpha }^{\pm }(m,r,T)=\sum_{n=0}^{\infty }(-1)^{n}\left[ K_{\alpha
}\left( \frac{nm}{T}+rm\right) \pm K_{\alpha }\left( \frac{(n+1)m}{T}%
-rm\right) \right] ,  \label{Kal}
\end{equation}
which will appear as a building block in the computation of many finite
temperature correlation functions. Since we intend to investigate the
commutativity of the diagram in figure 1, various limits of this function
will be required frequently. For $\func{Re}\alpha >0$ we compute with (\ref
{A2}) the massless limit

\begin{equation}
\lim_{m\rightarrow 0}\!\hat{K}_{\alpha }^{\pm }(m,r,T)\sim \frac{\Gamma
(\alpha )}{2^{1-\alpha }}\sum_{n=0}^{\infty }(-1)^{n}\left[ \left( \frac{nm}{%
T}+rm\right) ^{-\alpha }\pm \left( \frac{(n+1)m}{T}-rm\right) ^{-\alpha
}\right] \!\!\!.\!\!\!  \label{Kallim}
\end{equation}
The sum can be evaluated explicitly. We just report on the cases which are
important for our analysis 
\begin{eqnarray}
\lim_{m\rightarrow 0}\hat{K}_{1}^{+}(m,r,T) &\sim &\frac{\pi T}{m}\frac{1}{%
\sin (\pi rT)}\,,  \label{l+1} \\
\lim_{m\rightarrow 0}\hat{K}_{2}^{-}(m,r,T) &\sim &\left( \frac{\pi T}{m}%
\right) ^{2}\frac{2\cot (\pi rT)}{\sin ^{2}(\pi rT)}\,, \\
\lim_{m\rightarrow 0}\hat{K}_{3}^{+}(m,r,T) &\sim &\left( \frac{\pi T}{m}%
\right) ^{3}\frac{6+2\cos (2\pi rT)}{\sin ^{3}(\pi rT)}\,, \\
\lim_{m\rightarrow 0}m^{4}\hat{K}_{0}^{-}(m,r,T) &\sim &0\,, \\
\lim_{m\rightarrow 0}m^{4}\hat{K}_{0}^{+}(m,r,T)\hat{K}_{2}^{+}(m,r,T) &\sim
&0\,.
\end{eqnarray}
The zero temperature limit is more easily computed, just by noting that in
the sum of (\ref{Kal}) only the $n=0$ term survives 
\begin{equation}
\lim_{T\rightarrow 0}\hat{K}_{\alpha }^{\pm }(m,r,T)\sim K_{\alpha }\left(
rm\right) \,.  \label{KT0}
\end{equation}
A further function which frequently occurs is Lerch's transcendental
function, whose sum representation is 
\begin{equation}
\Phi (x,s,\alpha )=\sum_{n=0}^{\infty }\frac{x^{n}}{(n+\alpha )^{s}}\,\quad 
\text{for\quad }|x|<1,\,\alpha \notin \Bbb{Z}_{0}^{-}.  \label{Lerch}
\end{equation}
In many cases, however, we require precisely the value $x\rightarrow -1$ in (%
\ref{Lerch}). The convergence problem can be circumvented by exploiting the
fact that in the limit $x\rightarrow -1$ we can express $\Phi (x,s,\alpha )$
in terms of Riemann zeta functions 
\begin{equation}
\zeta (s,\alpha )=\sum_{n=0}^{\infty }\frac{1}{(n+\alpha )^{s}}\,\quad \text{%
for\quad }\limfunc{Re}s>1\,,
\end{equation}
instead 
\begin{equation}
\lim_{x\rightarrow -1}\Phi (x,s,\alpha )=\frac{1}{2^{s}}\left[ \zeta
(s,\alpha /2)-\zeta (s,(1+\alpha )/2)\right] \,.
\end{equation}
This leaves the problem $\lim_{x\rightarrow -1}\Phi (x,s,\alpha )$ for $%
\limfunc{Re}s<1$. Also the limit 
\begin{equation}
\lim_{\alpha \rightarrow 0}\Phi (-e^{-1/\alpha },s,\alpha )\sim \alpha
^{-s}\,\,,  \label{needed}
\end{equation}
is needed in section 3.


\begin{thebibliography}{99}
\bibitem{Kar}  {\small P.~Weisz, \emph{Phys. Lett.} \textbf{B67}, 179
(1977); M.~Karowski and P.~Weisz, \ \emph{Nucl. Phys.} \textbf{B139} (1978)
445.}

\bibitem{Smir}  {\small F.A.~Smirnov, \emph{Form factors in Completely
Integrable Models of Quantum Field Theory}, Adv. Series in Math. Phys. 
\textbf{14} (World Scientific, Singapore, 1992).}

\bibitem{massless}  {\small G.~Delfino, G.~Mussardo and P.~Simonetti, \emph{%
Phys. Rev.} \textbf{D51} (1995) R6621.}

\bibitem{Kirk}  {\small J.~Kirkwood, \emph{J. Chem. Phys}. \textbf{7} (1939)
39.}

\bibitem{Kubo}  {\small R.~Kubo, \emph{Can. J. Phys}. \textbf{34} (1956)
1274.}

\bibitem{LLSS}  {\small A.~LeClair, F.~Lesage, S.~Sachdev and H.~Saleur, 
\emph{Nucl. Phys. }\textbf{B482} [FS] (1996) 579.}

\bibitem{LM}  {\small A.~LeClair and G.~Mussardo, \emph{Nucl. Phys. }\textbf{%
B552} (1999) 624.}

\bibitem{HS}  {\small H.~Saleur, \emph{Nucl. Phys.} \textbf{B567} (2000) 602.%
}

\bibitem{GD}  {\small G.~Delfino, \emph{J. Phys. } \textbf{A34} (2001) L161.}

\bibitem{AW}  {\small G.~Mussardo, \emph{J. Phys. } \textbf{A34} (2001) 7399.%
}

\bibitem{FZ}  {\small L.D.~Faddeev, \emph{Sov. Sci. Rev. Math. Phys.} 
\textbf{C1} (1980) 107; A.B. Zamolodchikov and Al.B. Zamolodchikov, \emph{%
Ann. of Phys. } \textbf{120} (1979) 253.}

\bibitem{Balog}  {\small J.~Balog, \emph{Nucl. Phys. }\textbf{B419} (1994)
480.}

\bibitem{TBAZam}  {\small Al.B.~Zamolodchikov, \emph{Nucl. Phys.} \textbf{%
B358} (1991) 497.}

\bibitem{KMS}  {\small R.~Kubo, \emph{J. Math. Soc. Japan}. \textbf{12,}
(1957) 570; P.C.~Martin and J.~Schwinger, \emph{Phys. Rev. } \textbf{115}
(1959) 1342.}

\bibitem{Haag}  {\small R. Haag, \emph{Local Quantum Physics: Fields,
Particles, Algebras} 2-nd revised edition (Springer, Berlin, 1996).}

\bibitem{BPZ}  {\small A.A.~Belavin, A.M.~Polyakov and A.B.~Zamolodchikov, 
\emph{Nucl. Phys.} \textbf{B241} (1984) 333. }

\bibitem{Buch}  {\small D.~Buchholz, \emph{Lect. Notes Phys.} \textbf{539}
(2000) 211. }

\bibitem{Todo}  {\small I.T.~Todorov, \emph{``Finite Temperature
2-Dimensional QFT Models of Conformal Current Algebra''} in \emph{%
``Conformal Invariance and String Theory''} ed. P.~Dita and V.~Georgescu,
(Academic Press, London, 1989); D.~Buchholz, G. Mack and I.T.~Todorov, \emph{%
Nucl. Phys. }\textbf{B5 } [Proc. Suppl.] (1988) 20.}

\bibitem{triZam}  {\small Al.B.~Zamolodchikov,\emph{\ Nucl. Phys.} \textbf{%
B358} (1991) 524.}

\bibitem{Hecht}  {\small R.~Hecht, \emph{Phys. Rev.} \textbf{158} (1967) 557.%
}

\bibitem{DSC}  {\small G.~Delfino, P.~Simonetti and J.L.~Cardy, \emph{Phys.
Lett}. \textbf{B387} (1996) 327.}

\bibitem{Feder}  {\small P.~Federbush, \emph{Phys. Rev.} \textbf{121} (1961)
1247; \emph{Progress of Theo. Phys.} \textbf{26} (1961) 148.}

\bibitem{Fform}  {\small O.A.~Castro-Alvaredo and A.~Fring, \emph{Nucl.
Phys. }\textbf{B618} [FS] (2001) 437.}

\bibitem{book}  {\small P.~Di Francesco, P.~Mathieu and D.~S\'{e}n\'{e}chal, 
\emph{Conformal Field Theory } (Springer, New York, 1997).}

\bibitem{CM}  {\small J.~Cardy and G.~Mussado, \emph{Phys. Lett.} \textbf{%
B225} (1989) 275.}

\bibitem{FFZam}  {\small Al.B.~Zamolodchikov, \emph{Nucl. Phys.} \textbf{B348%
} (1991) 619.}

\bibitem{OAnew}  {\small O.A.~Castro-Alvaredo and A.~Fring, in preparation.}

\bibitem{OAF}  {\small O.A.~Castro-Alvaredo, A.~Fring and F.~G\"{o}hmann, 
\emph{``On the absence of simultaneous reflection and transmission in
integrable impurity systems''}, hep-th/0201142.}

\bibitem{SL}  {\small S.~Lukyanov, \emph{Nucl. Phys. } \textbf{B612} (2001)
391.}
\end{thebibliography}
\end{document}